\pdfoutput=1

\documentclass[11pt]{article}

\usepackage{authblk}

\usepackage[preprint]{acl}

\usepackage{times}
\usepackage{latexsym}

\usepackage[T1]{fontenc}

\usepackage[utf8]{inputenc}

\usepackage{microtype}

\usepackage{inconsolata}
\usepackage{multirow}

\usepackage{graphicx}

\usepackage{amsmath}
\usepackage{amssymb}

\usepackage{graphicx}
\usepackage{caption}
\usepackage{subcaption}
\usepackage{float} 
\usepackage{hyperref}
\usepackage{enumitem}

\usepackage{xcolor}
\usepackage{soul}
\usepackage{booktabs}
\usepackage[normalem]{ulem}
\useunder{\uline}{\ul}{}

\title{Multimodal Language Models with Modality-Specific Experts for Financial Forecasting from Interleaved Sequences of Text and Time Series}

\author[1,3]{Ross Koval}
\author[2]{Nicholas Andrews}
\author[1]{Xifeng Yan}
\affil[1]{University of California, Santa Barbara}
\affil[2]{Johns Hopkins University} 
\affil[3]{AJO Vista} 
\affil[ ]{\texttt {rkoval@ucsb.edu}}

\begin{document}
\maketitle
\begin{abstract}
Text and time series data offer complementary views of financial markets: news articles provide narrative context about company events, while stock prices reflect how markets react to those events. However, despite their complementary nature, effectively integrating these interleaved modalities for improved forecasting remains challenging. In this work, we propose a unified neural architecture that models these interleaved sequences using modality-specific experts, allowing the model to learn unique time series patterns, while still enabling joint reasoning across modalities and preserving pretrained language understanding capabilities. To further improve multimodal understanding, we introduce a cross-modal alignment framework with a salient token weighting mechanism that learns to align representations across modalities with a focus on the most informative tokens. We demonstrate the effectiveness of our approach on a large-scale financial forecasting task, achieving state-of-the-art performance across a wide variety of strong unimodal and multimodal baselines. We develop an interpretability method that reveals insights into the value of time series-context and reinforces the design of our cross-modal alignment objective. Finally, we demonstrate that these improvements translate to meaningful economic gains in investment simulations.
\end{abstract}

\section{Introduction}
Text and time series provide complementary perspectives on financial markets. News articles describe company events, such as earnings announcements, product launches, and mergers, while stock prices reflect how markets react to these events over time. When these modalities are combined, it presents a promising yet challenging opportunity for large language models (LLMs) to enhance financial forecasting by reasoning over temporally aligned but semantically distinct inputs. 

\begin{figure}[t]
\centering
\scalebox{0.99}{
\includegraphics[width=0.5\textwidth]{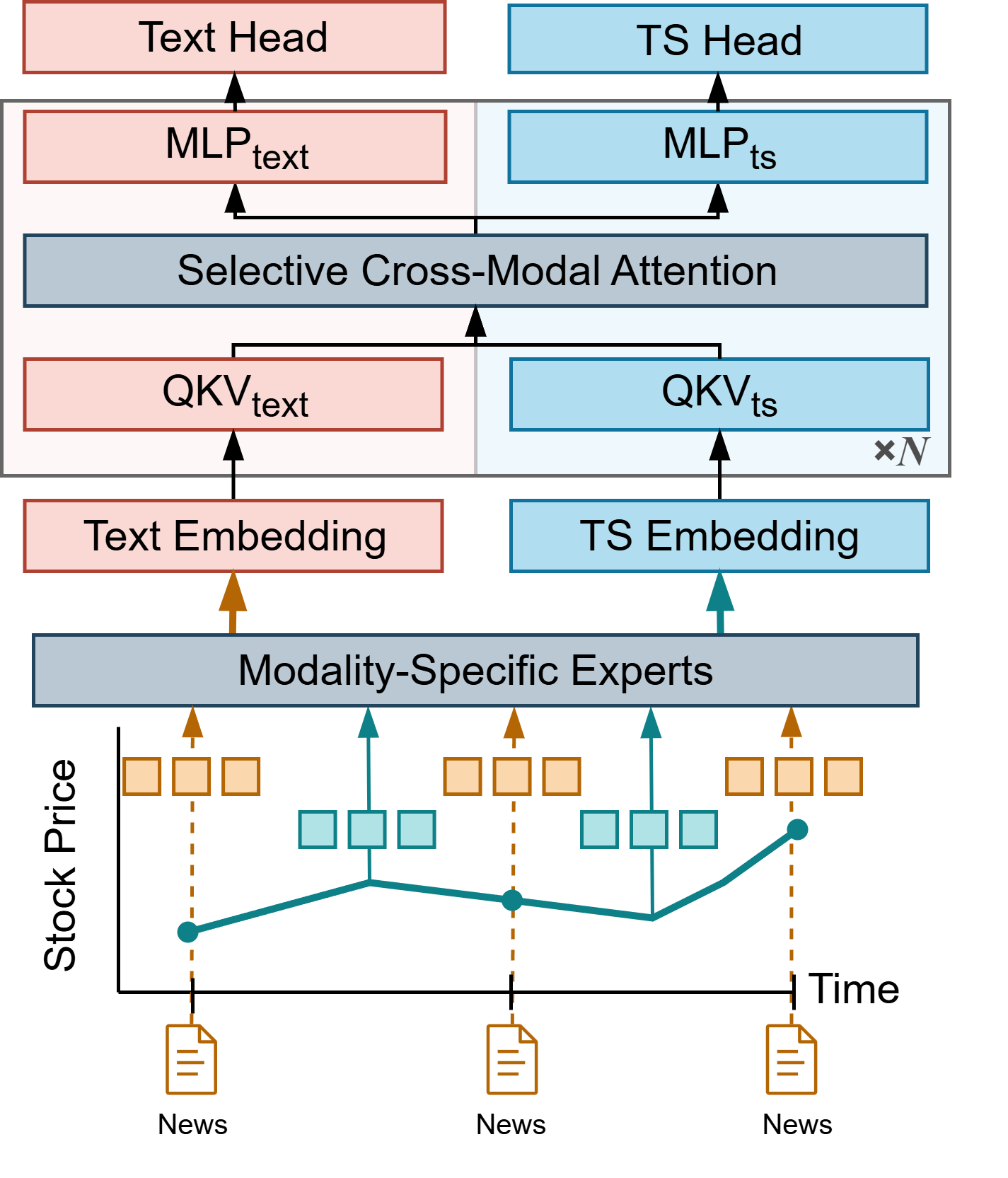}
}
\caption{Overview of our multimodal forecasting task and proposed model (MSE-ITT), which processes interleaved sequences of tokens of news articles (Text) and discretized stock returns (TS). MSE-ITT incorporates modality-specific experts to capture distinct patterns in text and time series, while enabling joint reasoning across modalities through selective cross-modal attention.}
\label{fig:model_diagram}
\end{figure}

To better understand their complementary nature, consider how time series and text contribute distinct but related signals. The time series data offers global context into both short and long-term price behavior, allowing the model to learn patterns driven by investor biases \citep{jegadeesh1990evidence, jegadeesh1993returns, kelly2021understanding}. Moreover, the interleaved stock prices act as implicit supervision, revealing how markets have previously responded to similar news, and exposing the model to repeated cause-and-effect dynamics. Conversely, news articles provide both retrospective context about historical market behavior and forward-looking signals that may reinforce or contradict current price trends. By jointly modeling these modalities, LLMs have the potential to learn cross-modal interactions and context-aware representations that would not be possible from either modality alone.

However, effectively integrating long sequences of text and time series in a unified model remains challenging. Simple strategies, such as converting numerical data into strings of digits, fail to capture their distinct structures. Language is discrete, syntax-rich, and compositional, while time series are continuous, stochastic, and governed by temporal dependencies. Furthermore, news arrives at irregular intervals, while stock prices are observed daily. These structural differences make it difficult for pretrained LLMs, which are optimized for language, to extract meaningful signals from time series inputs. Instead, we argue that modeling these multimodal sequences requires modality-specific components that can respect and exploit the unique structure of each input type, while enabling joint reasoning across modalities.

To address this, we propose a unified multimodal architecture with modality-specific experts and design a pretraining objective to effectively learn cross-modal interactions. In summary, we make the following key contributions.

\begin{enumerate}[leftmargin=*]
  \item We design a multimodal architecture (\textbf{MSE-ITT}) that can effectively model interleaved sequences of text and time series, demonstrating state-of-the-art performance on a challenging financial forecasting task compared to a comprehensive set of baselines (\autoref{sec:methods}, \autoref{tab:main_results}).
  \item We introduce a unified cross-modal alignment framework (\textbf{SALMON}) with a dynamic, salient token weighting mechanism (\textbf{STW}) to effectively learn time-series-specific features that enhance language understanding (\autoref{sec:cmlm}, \autoref{tab:ablations_pretraining}).
  \item We develop an interpretability method that reveals insights into the value of time series context and demonstrate that it translates to significant economic gains in investment applications (\autoref{sec:value_ts_context}, \autoref{tab:lm_dict}, \autoref{tab:port_sims}).
\end{enumerate}
\paragraph{Broader Impact} We hope this work encourages future research on LLMs that reason over interleaved sequences of text and time series, particularly in domains where structured and unstructured data interact over time, such as finance, healthcare, and climate. Our findings highlight the limitations of treating time series equivalently to language and underscore the importance of dedicated mechanisms for structured time series inputs. To support this research direction, we release the code at: \url{https://github.com/rosskoval/mlm_text_ts}.

\section{Related Work}
\subsection{Multimodal Time Series Forecasting}
LLMs have greatly improved their capabilities in understanding multimodal inputs, such as text, images, audio, and video \citep{liu2023visual,li2024audio, team2024chameleon, chen2023videollm}. However, they have demonstrated challenges in understanding time series data. While some work has found benefit in using LLMs to perform time series forecasting with zero-shot learning \citep{gruver2023large} or finetuning \citep{jin2023time}, other work has found that their pretrained weights do not provide positive transfer \citep{Tan2024LLMsUseful} and struggle to reason about them effectively \citep{Merrill2024LLMsStruggle} without specialized encoding \citep{Chow2024TimeSeriesReasoning}. Moreover, recent work on contextualized forecasting has explored integrating text and time-series data using two main strategies. One paradigm converts time series into strings or embeddings, allowing text to condition LLM predictions \citep{jin2023time, Williams2024ContextIsKey, Kim2024MMF}. However, this assumes LLMs can natively interpret time series and limits the model’s ability to learn modality-specific patterns. Another approach uses frozen language models to extract fixed text features for multivariate time series models \citep{Liu2024TimeMMD, Li2025TaTS}. While this allows modality-specific patterns, it prevents deep cross-modal interaction and inhibits the reasoning capabilities of LLMs. 

\subsection{Financial Prediction}
Recent work has shown that language models can predict stock returns from news articles \citep{chen2022expected, xie2023pixiu, lopez2023can}. Separately, historical price patterns have been shown to be predictive \citep{jegadeesh1990evidence,jegadeesh1993returns,kelly2021understanding}. These findings have motivated recent work exploring methods to integrate textual and time series data to improve forecasting \citep{xu2018stock, ang2022guided, koval-etal-2024-financial, Wang2024ModelingNews, zong2024stock, mou2025mm}. However, most of these models process each modality independently and generally rely on late-fusion methods, limiting cross-modal interaction during representation learning. 

\section{Problem Statement}\label{sec:problem}

\subsection{Multimodal Inputs}\label{multimodal_inputs}
In our main problem formulation, we consider a multimodal sequence of time-stamped inputs, aperiodically arriving news articles (text), and daily stock returns (time series), described below. 

\paragraph{Time Series Inputs}\label{ts_inputs}
As time series inputs $X_t$, we consider the daily stock return $r_t$ of the company over the last 1-year (252 trading days). 
$$X_t = \{r_{t-252},..., r_{t-1}\} $$
While we demonstrate that our method generalizes to multivariate time series in \autoref{sec:appendix:multivariate}, we focus on the univariate case in our main experiments to more precisely study the effects of the cross-modal interaction method in a controlled setting.

\paragraph{Textual Inputs}\label{text_inputs}
As textual inputs $Y_t$, we consider the $N$ most recent news articles $a_t$ about the company prior to time $t$:
$$Y_t = \{a_{t-N},..., a_{t-1}\}$$

For computational efficiency, we select the $N=10$ most recent articles about the same company that occurred within the last 1-year.

\subsection{Task Formulation}\label{sec:formulation}
Following \citet{xu2018stock, chen2022expected}, we adopt the task of predicting the direction of the stock price $P_t$ change over the course of short-term and long-term horizons (days) $h\in \{\mathrm{7D}, \mathrm{30D}\}$ at prediction time $t$.
$$ D_{t} = \mathrm{Sign}(P_{t+h} - P_{t}) $$
We evaluate the performance on this binary classification task with AUC because the continuous scores are more informative than discrete classes for investment management applications.

\paragraph{Data Acquisition and Curation}
We curate financial news articles in English from the \href{https://huggingface.co/datasets/Zihan1004/FNSPID}{FNSPID Dataset} \citep{dong2024fnspid}, for US-based public companies, which covers a variety of company events and news sources. We perform a curation process, following the filtering criterion in \citet{chen2022expected} for data quality, detailed in \autoref{sec:appendix:data_curation}. Our sample contains more than 3,000 public companies in the US, encompassing a diverse range of firm sizes and industries. 
 
\paragraph{Data Statistics and Task Formulation}
We temporally partition the data into training (2010-2017), validation (2018-2019), and test (2020-2024) sets. We provide summary statistics in~\autoref{tab:stats}.

\begin{table}[htp]
\centering
{\small
\scalebox{0.90}{%
\begin{tabular}{@{}lrrr@{}}
\toprule
\multicolumn{1}{c}{{\ul }}                                                     & \textbf{Train} & \textbf{Validation} & \textbf{Test} \\ \midrule
Start Date                                                                     & Jan-2010         & Jan-2018              & Jan-2020     \\ \midrule
End Date                                                                       & Dec-2017         & Dec-2019              & Dec-2024     \\ \midrule
\# Samples                                                                     & 155,146           & 36,931                & 115,611        \\ \midrule
\begin{tabular}[c]{@{}l@{}} \# Companies \end{tabular}                 & 2,591          & 2,249               & 3,564             \\ \bottomrule
\end{tabular}
}
\caption{Summary statistics of the characteristics of financial news articles and stock return time series in each sample split.}
\label{tab:stats}
}
\end{table}
\section{Proposed Method}\label{sec:methods}
\subsection{Model Design}
In this section, we introduce our proposed modality-specific experts multimodal model for interleaved sequences of text and time series (\textbf{MSE-ITT}), illustrated in \autoref{fig:model_diagram}. The use of mixture-of-experts (MOE) layers \citep{shazeer2017outrageously, fedus2022switch, zoph2022st} in LLMs have become pervasive because of their improved representational capacity and compute efficiency, allowing subnetworks to specialize in different regions of the input. Inspired by recent advances in multimodal MOE-based LMs for text and images \citep{Zhou2024Transfusion,Shi2024LMFusion,Lin2024MoMa,Liang2025MixtureMamba}, we design a modality-specific experts architecture that builds on Llama3-8B \citep{llama3_herd_2024}, an autoregressive LM with strong language capabilities, and extend it with components tailored for temporal and cross-modal understanding. 

\paragraph{Modality-Specific Experts} To process both modalities effectively, we introduce dedicated modality-specific (TS) parameters into each layer of the pretrained LM. These modality-specific components are responsible for processing time series inputs, enabling the model to capture time-series patterns without disrupting the pretrained language capabilities of the base LM. This separation reduces cross-modal interference and imposes an inductive bias that respects the structural differences between text and time series. At the same time, it allows for joint modeling of interleaved sequences. We maintain causal masking in the self-attention layers to preserve temporal causality.

These added parameters include layer normalization ($\mathrm{LN}$), multi-headed attention projections ($\mathrm{QKV}$), and feedforward MLPs ($\mathrm{MLP}$):
\[
\mathbf{h}_{\text{text}}^{q},\,
\mathbf{h}_{\text{text}}^{k},\,
\mathbf{h}_{\text{text}}^{v}
=\mathrm{QKV}_{\text{text}}\!\bigl(\mathrm{LN}_{\text{text}}(\mathbf{h}_{\text{text}})\bigr)
\]
\[
\mathbf{h}_{\text{ts}}^{q},\,
\mathbf{h}_{\text{ts}}^{k},\,
\mathbf{h}_{\text{ts}}^{v}
=\mathrm{QKV}_{\text{ts}}\!\bigl(\mathrm{LN}_{\text{ts}}(\mathbf{h}_{\text{ts}})\bigr)
\]
\[
\mathbf{h}_{\text{text}} =
\mathrm{MLP}_{\text{text}}\!\bigl(\mathrm{LN}_{\text{text}}(\mathbf{h}_{\text{text}})\bigr)
\]
\[
\mathbf{h}_{\text{ts}} =
\mathrm{MLP}_{\text{ts}}\!\bigl(\mathrm{LN}_{\text{ts}}(\mathbf{h}_{\text{ts}})\bigr)
\]

In this design, the text ($\mathbf{h}_{\text{text}}$) and time series ($\mathbf{h}_{\text{ts}}$) hidden states are routed to separate modules, before performing joint self-attention over the multimodal sequence:
\[
\mathbf{h}=
\operatorname{softmax}\!\Bigl(
  [\mathbf{h}_{\text{text}}^{q};\mathbf{h}_{\text{ts}}^{q}]
  [\mathbf{h}_{\text{text}}^{k};\mathbf{h}_{\text{ts}}^{k}]^{\!\top}
\Bigr)
\,[\mathbf{h}_{\text{text}}^{v};\mathbf{h}_{\text{ts}}^{v}]
\]

We initialize these dedicated TS-parameters from the pretrained values in the corresponding LM layer. We omit the residual connections and scaling for brevity. 

\paragraph{Selective Cross-Modal Attention}
Prior work has shown that early layers of pretrained LMs capture low-level, syntactic patterns \citep{clark-etal-2019-bert,nagrani2021attention,zhang2025unravelling}, while deeper layers encode higher-level, global semantics. In addition, attention weights have been found to exhibit noise due to positional biases \citep{liu2023lost,xiao2023efficient,ye2024differential,golovneva2025multi}. Based on these findings, our hypothesis is that deeper layers are more likely to benefit from time-series context and that introducing cross-modality attention at early layers may be harmful, as it risks disrupting low-level, modality-specific representations with noisy time-series signals. To address this, we restrict cross-modality attention to the top half of the network layers (16-32), allowing the lower layers (1-16) to focus on learning more localized, modality-specific features. This design choice results in efficiency gains, and, as shown in our ablation studies (\autoref{tab:ablations_architecture}), improves language understanding and task performance.

\paragraph{Interleaved Multimodal Sequence}
We design the input as a temporally ordered sequence of multimodal tokens, where $a_{t}$ is a sequence of tokens for news article $a$ and $r_t$ is the stock return with timestamp $t$. 
$$ x_{1:L} = \{r_{t-252},...,a_{t-N},...,r_{t-1},...,a_{t-1}\} $$
Since the news articles arrive at irregular intervals, we employ pointwise embedding tokenization \citep{shi2024time} to accommodate a variable number of time steps between news articles, rather than using fixed-length patches. This allows flexible modeling of sequences with variable temporal resolution and supports application to other domains with irregular event streams. 

Because financial data is known to be statistically noisy, we discretize the continuous time series values into a fixed number of bins $B$, enabling more robust representations and reducing sensitivity to outliers. We learn embeddings ($\text{TSEmb}$) for each bin \citep{rabanser2020effectiveness, ansari2024chronos} with quantile binning to ensure a well-calibrated distribution. We tune $B$ over the validation set and ablate this design choice in \autoref{sec:appendix:ts_quant}. This choice allows us to transform the inputs into a unified sequence of embeddings $z_i$:
\begin{equation*}
\mathbf{z}_i =
\begin{cases}
\text{TextEmb}\big(x_i\big), & \text{if } \text{mod}(x_i) = \text{text}, \\[4pt]
\text{TSEmb}\big(x_i\big),    & \text{if } \text{mod}(x_i) = \text{ts},
\end{cases}
\label{eq:input}
\end{equation*}
We pass them through MSE-ITT to obtain contextualized hidden states $\mathbf{h}$:
\begin{equation*}
\mathbf{h}_{1:L} = \text{MSE-ITT}\!\bigl(\mathbf{z}_{1:L}\bigr)
                 = [\,\mathbf{h}_1,\dots,\mathbf{h}_L\,]
\label{eq:hidden}
\end{equation*}
This interleaved input design enables the model to leverage the pretrained LLM’s relative positional encodings \citep{su2024roformer}, to reflect the natural temporal order of events, rather than relying on learnable time embeddings \citep{woo2024unified} only at the input layer. Because rotary encodings impose a strong locality bias throughout the network, preserving temporal order in the input allows the model to better exploit this inductive bias in its attention patterns. It also ensures that the relative distance between article tokens reflects their actual separation in time, spaced by time series embeddings corresponding to the number of days between events.

\subsection{SALMON}\label{sec:cmlm}
To further improve multimodal understanding, we introduce Salience-Aware Language Modeling over Interleaved Modalities (\textbf{SALMON}), a unified pretraining objective, that jointly predicts the next text token and the next time series token in an interleaved multimodal sequence, with a dynamic salience-based weighting mechanism. Since we discretize the time series into discrete tokens, both objectives can be trained using cross-entropy loss, with separate projection heads for text tokens ($U_{\text{text}}$) and time series tokens ($U_{\text{ts}}$):

\begingroup\small
\begin{equation*}
P_\theta(x_{i})=
\begin{cases}
\operatorname{softmax}(U_{\text{text}}\mathbf{h}_{i-1}), & \text{if } \mathrm{mod}(x_{i})=\text{text},\\
\operatorname{softmax}(U_{\text{ts}}\mathbf{h}_{i-1}),   & \text{if } \mathrm{mod}(x_{i})=\text{ts},
\end{cases}
\end{equation*}
\endgroup

\begingroup\small
\begin{equation*}
\mathcal{L}(x;\theta)
= - \sum_{i\in\mathcal{I}_{\text{text}}}\!
\log P_{\theta}(x_{i,\text{text}})
\;-\;
\sum_{i\in\mathcal{I}_{\text{ts}}}\!
\log P_{\theta}(x_{i,\text{ts}})
\end{equation*}
\endgroup

To enable cross-modal learning without disrupting pretrained language capabilities, we freeze the pretrained text-branch parameters and train only the newly introduced TS-specific parameters. This joint objective encourages the model to learn to identify features in the time series that are predictive of future news text, and learn aligned representations that capture how text and time series data co-evolve together. 
\begin{table*}[t]
\centering
{\small
\scalebox{0.95}{%
\begin{tabular}{@{}llllcc@{}}
\toprule
\textbf{Model Class} & \textbf{Model} & \textbf{Base LM} & \textbf{Input} & \textbf{7D} & \textbf{30D} \\ \midrule
\multirow{6}{*}{\textit{Zero-Shot LLM}} 
  & GPT-4o, Direct \citep{lopez2023can} & GPT-4o & text & 52.45 & 53.31 \\
  & GPT-4o, CoT & GPT-4o & text & 53.15 & 54.33 \\
  & GPT-4o, Direct & GPT-4o & ts & 50.92 & 49.77 \\
    & GPT-4o, CoT & GPT-4o & ts & 48.70 & 47.90 \\
  & GPT-4o, Direct \citep{Williams2024ContextIsKey} & GPT-4o & text, ts & 52.09 & 52.71 \\
  & GPT-4o, CoT \citep{Tan2025InferringEvents} & GPT-4o & text, ts & 50.56 & 53.05 \\ \midrule
\multirow{2}{*}{\textit{Unimodal}} 
  & TS-Only \citep{nie2022time} & None & ts & 52.93 & 53.88 \\
  & Text-Only \citep{chen2022expected} & Llama3-8B & text & 53.76 & 54.13 \\ \midrule
\multirow{5}{*}{\textit{MMTSF}} 
  & TimeLLM \citep{jin2023time} & Llama2-7B & text, ts & 53.79 & 55.10 \\
  & TaTs \citep{Li2025TaTS} & Llama2-7B & text, ts & 54.48 & 54.81 \\
  & TTSR \citep{Chow2024TimeSeriesReasoning} & Mistral-7B & text, ts & 55.93 & 56.17** \\
  & TimeMDD \citep{Liu2024TimeMMD} & Llama3-8B & text, ts & 55.15 & 55.25 \\
  & Hybrid-MMF \citep{Kim2024MMF} & Llama3-8B & text, ts & 55.96* & 55.84 \\ \midrule
\multirow{6}{*}{\textit{SFF}} 
  & FinMA \citep{xie2023pixiu} & Llama2-7B & text, ts & 51.11 & 52.15 \\
  & MTFE-MICM \citep{koval-etal-2024-financial} & BigBird-125M & text, ts & 55.44 & 54.49 \\
  & FININ \citep{Wang2024ModelingNews} & RoBERTa-125M & text, ts & 52.47 & 53.13 \\
  & StockTime \citep{wang2024stocktime} & Llama3-8B & text, ts & 55.36 & 55.85 \\
  & MAT \citep{emami2024modality} & FinBERT-110M & text, ts & 54.43 & 53.81 \\
  & MM-iTransformer \citep{mou2025mm} & FinBERT-110M & text, ts & 54.16 & 53.57 \\ \midrule
\textbf{Proposed} 
  & \textbf{MSE-ITT} & Llama3-8B & text, ts & \textbf{57.94}* [0.08] & \textbf{58.48}** [0.07] \\ \bottomrule
\end{tabular}%
}}
\caption{Main Results (higher is better): Model performance on the test set of our multimodal prediction task at different forecasting horizons. All results indicate the AUC of the model's predicted probabilities reported in percentage units. "[]" indicate the sample standard deviation of the results over 3 training runs with different random seeds. *, ** indicate that the performance of our proposed model is statistically better ($p<0.01$) than the next best performing model according to DeLong's test. The last row indicates our proposed method MSE-ITT.}
\label{tab:main_results}
\end{table*}

\begin{figure}[t]
\centering
\scalebox{0.85}{
\includegraphics[width=0.5\textwidth]{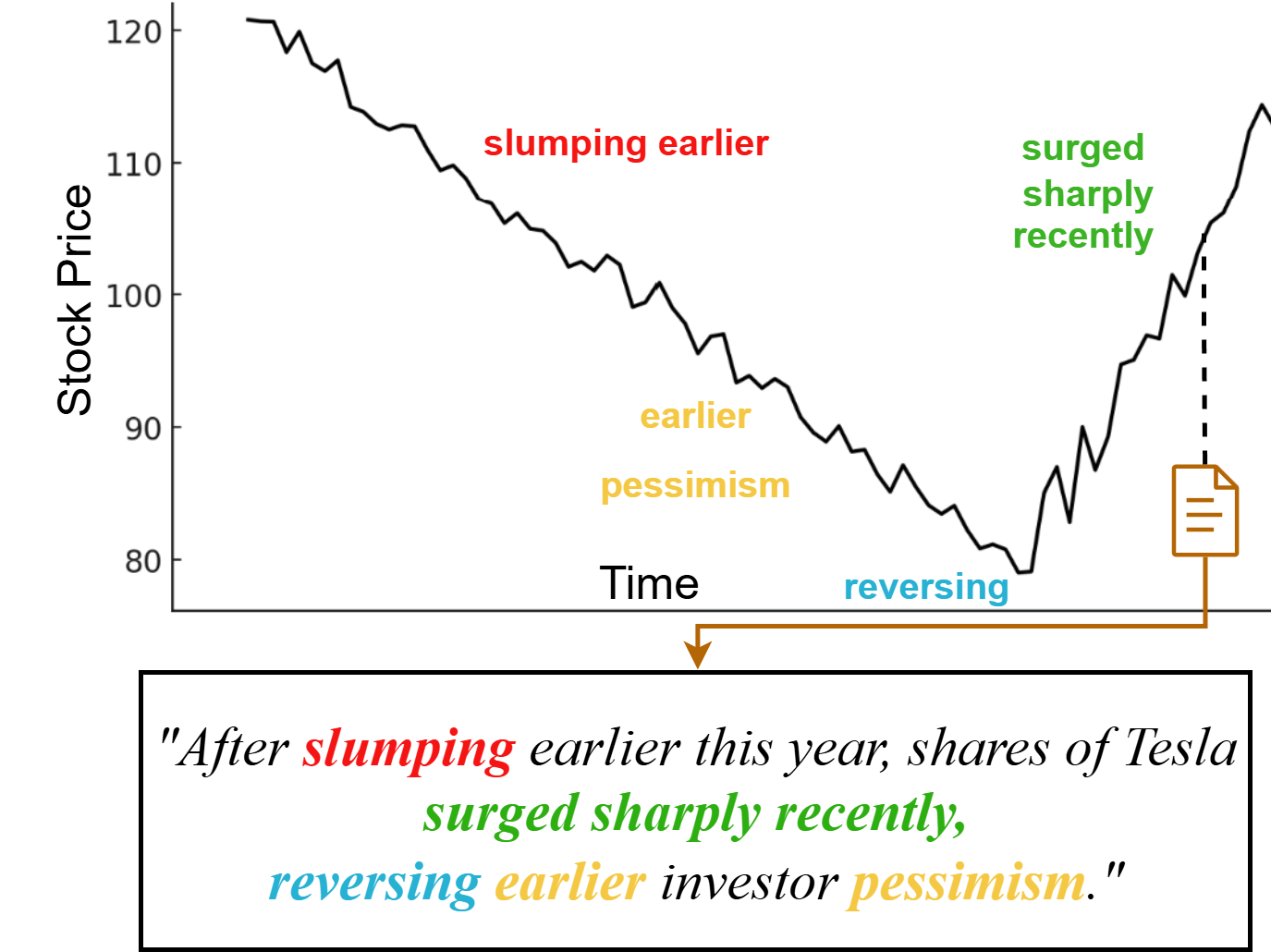}
}
\caption{Illustration of our cross-modal alignment framework, SALMON, which learns to align historical stock price behavior and news articles with a unified objective. The Salient Token Weighting (STW) mechanism dynamically assigns higher weight to tokens that benefit most from time-series context, improving cross-modal alignment.}
\label{fig:cma_diagram}
\end{figure}

\paragraph{Salient Token Weighting (STW)}\label{sec:salient_pretraining}
While SALMON enables joint language modeling across modalities, standard cross-entropy loss places equal weight on all next token predictions. However, we hypothesize that not all textual tokens should theoretically or equally benefit from time-series context. In typical news text, many tokens, such as function or filler words, are easily predicted using neighboring tokens. While these tokens may not benefit from TS-context, we suspect that specific salient tokens, such as those sentiment-charged or related to market behavior, could benefit substantially. For example, consider the news headline in \autoref{fig:cma_diagram}.

We hypothesize that the easily predicted words dominate the loss function and that a selective weighting mechanism can help focus the training signal on tokens where TS-context provides significant mutual information, thereby improving the alignment process by reducing weight on noisy, irrelevant tokens.

To systematically identify such salient tokens without relying on external sentiment dictionaries, we design a contrastive estimation approach inspired by recent advances in contrastive decoding \citep{li-etal-2023-contrastive,yuan-etal-2024-speculative} and long-context training \citep{ye2024differential, fang2024wrong}. To this end, we leverage the LM with text-only inputs as a contrastive baseline. Specifically, we compute two versions of token-level predicted probabilities: one using text-only inputs and another using the full multimodal input. Since we freeze the text-only parameters of the model, we can compute the baseline forward pass efficiently with no gradients attached. The ratio of these probabilities (i.e. the likelihood ratio) reflects how much the TS context improves the prediction of each token, and serves as a proxy for token salience.

Mathematically, assume that \( P_{\theta}(x_{i,\text{text}} \mid x_{j < i,\text{text}}) \) is the probability of the \( i \)-th text token given preceding text tokens, and \( P_{\theta}(x_{i,\text{text}} \mid x_{j < i,\text{text}}, x_{j < i,\text{ts}}) \) is the probability when preceding time series context is also provided. We define the token-level weight as:
\begin{equation*}
W(x_{i,\text{text}}) = \frac{P_{\theta}(x_{i,\text{text}} \mid x_{j < i,\text{text}}, x_{j < i,\text{ts}})}{P_{\theta}(x_{i,\text{text}} \mid x_{j < i,\text{text}})}
\end{equation*}
such that $W$ directly measures how much the prediction of each token improves when time series context is available. Values greater than 1 indicate tokens that benefit from time series context, while values less than 1 indicate tokens where time series context is less helpful or potentially distracting. Finally, the weights are normalized to have mean one per sequence $\tilde{W}$ and then applied to each textual token in the cross-entropy loss. Mathematically, let $x_{i,\text{text}}$ and $x_{i,\text{ts}}$ denote the $i$-th text and time series tokens, respectively, and $c_{j<i} = (x_{j<i,\text{text}}, x_{j<i,\text{ts}})$ denote the previous context, then the $STW$ loss is formed on textual inputs $x_{\text{text}}$ by:
$$
\begin{aligned}
\mathcal{L}_{STW}(x_\text{text}; \theta) ={}& -\sum_{i\in\mathcal{I}_{\text{text}}} \tilde{W}(x_{i,\text{text}}) \cdot \\
& \log P_{\theta}(x_{i,\text{text}} \mid c_{j<i})
\end{aligned}
$$
This weighting mechanism amplifies the learning signal for textual tokens that benefit most from time series context. Since, at the beginning of training, the estimated token-level weights are not meaningful as the model has not yet learned to effectively leverage the time series context, we warm-start $W = 1$ for the first 20\% of training steps, and then relax this constraint to $W\in[0.1, 10.0]$ as the models learn to better utilize this information. We perform this cross-modal pretraining step on the input training data, and ablate the value of it and the token weighting mechanism in \autoref{tab:ablations_pretraining}.

\paragraph{Implementation Details}
Our method is parameter and compute efficient as we freeze the pretrained text-branch parameters, and only finetune the newly added TS-branch parameters efficiently with LoRA \citep{hu2021lora}, during both cross-modal alignment and task finetuning. 

\section{Baselines}\label{sec:baselines}
We provide a comprehensive set of strong baselines to evaluate the benefits of our approach, spanning commercial LLMs, state-of-the-art multimodal time series forecasting models, and specialized financial models for stock movement prediction. 

\paragraph{Unimodal}\label{sec:unimodal_baselines}
First, we provide simple unimodal baselines that indicate the independent forecasting ability of each input modality. \textbf{Text-Only} indicates training a classifier on frozen Llama3-8B embeddings of the text-only inputs \citep{ke2019predicting}. \textbf{TS-Only} indicates training PatchTST \citep{nie2022time} on only the time series inputs. 

\paragraph{Zero-Shot LLMs}\label{sec:llm_baselines}
We include the zero-shot performance of GPT-4o \citep{openai2024gpt4o} with a variety of different prompting configurations, following \citep{Williams2024ContextIsKey}, establishing state-of-the-art commercial LLM baselines and highlighting the difficulty of the task. We provide different prompting methods, including direct prediction (\textbf{Direct}) and Chain-of-Thought (\textbf{CoT}) \citep{kim2024financial,Tan2025InferringEvents}. The time series inputs are converted to text strings and we tune their formatting with validation set performance \citep{gruver2023large, Williams2024ContextIsKey}. We also provide text-only and time-series only baselines to indicate the relative ability of LLMs to reason over each modality, detailed in \autoref{sec:appendix:zeroshot_llms}. 

\paragraph{Multimodal Time Series Baselines}\label{sec:multimodal_baselines}
We implement a variety of baselines specialized for multimodal time series forecasting (\textbf{MMTSF}). These include TaTs \citep{Li2025TaTS}, TimeMDD \citep{Liu2024TimeMMD}, TimeLLM \citep{jin2023time}, Hybrid-MMF \citep{Kim2024MMF}, and TTSR \citep{Chow2024TimeSeriesReasoning}. All of these models are finetuned on the training data according to their original implementations, described further in \autoref{sec:appendix:baseline_models}.

\paragraph{Financial Forecasting Baselines}\label{sec:ff_baselines}
We include a variety of baselines specialized for multimodal stock movement prediction (\textbf{SFF}). These include FinMA \citep{xie2023pixiu}, MTFE-MICM \citep{koval-etal-2024-financial}, FININ \citep{Wang2024ModelingNews}, StockTime \citep{wang2024stocktime}, MAT \citep{emami2024modality}, and MM-iTransformer \citep{mou2025mm}.  All of these models are finetuned on the training data, detailed in \autoref{sec:appendix:baseline_models}.

\section{Experimental Results and Analysis}\label{sec:results}

\subsection{Main Results}\label{sec:main_results}
Overall, we find that our proposed model delivers meaningful gains in forecasting performance across time horizons compared to a strong set of diverse baselines, and that these gains translate to significant improvements in investment simulations (\autoref{tab:port_sims}). As we further demonstrate with ablation experiments (\autoref{tab:ablations_pretraining}, \autoref{tab:ablations_architecture}), these improvements stem from key architectural design choices grounded in inductive biases.  First, the model captures modality-specific structure by introducing dedicated parameters for time series data, while maintaining a unified architecture that enables joint reasoning across modalities. Second, our SALMON objective aligns time series features in the latent space of the LLM for enhanced cross-modal understanding, while our dynamic salient weighting mechanism enhances cross-modal alignment by focusing on key tokens that benefit most from time series context.

\paragraph{Time Series Context} The main results in \autoref{tab:main_results} highlight the challenging nature of the task and that the value of multimodal context highly depends upon the method of integration. While some methods benefit significantly from multimodal context, others underperform unimodal baselines.

We highlight that across both general and specialized baselines, there is a clear trend that methods that initially encode the time series with separate parameters from the LLM perform better than those that exclusively rely on LLM's native ability to extract meaningful time series features in either the text or latent space. Similarly, we find that models that simply treat text features as additional channels within a multivariate time series model underperform those that harness the powerful reasoning capabilities of LMs. These findings reinforce our design decision for modality-specific experts within a unified architecture. 

\paragraph{Zero-Shot LLMs} We highlight two interesting findings for GPT-4o. Firstly, GPT-4o performs better with text-only inputs than multimodal inputs, failing to benefit from the time series context. Secondly, prompting with chain-of-thought improves the ability to GPT-4o to analyze the textual inputs while failing to improve its ability to reason about the time series inputs. Overall, these results clearly highlight the challenges that LLMs exhibit in reasoning over time series inputs \citep{Merrill2024LLMsStruggle}, particularly in the financial domain, which are often noisy and lack consistent patterns.

\begin{table}[htp]
\centering
{\small
\scalebox{1.0}{%
\begin{tabular}{@{}ccc@{}}
\toprule
\textbf{Method} & \textbf{7D} & \textbf{30D} \\ \midrule
MSE-ITT w/o SALMON            &  56.93    &    57.14              \\
+SALMON w/o STW           &  57.56     &    57.89              \\
+SALMON w/ STW            &  57.94     &    58.48              \\
\bottomrule
\end{tabular}
}
\caption{Results demonstrate the value of our proposed Cross-Modal  alignment objective (SALMON) with Salient-Token Weighting (STW) compared to the baseline MSE-ITT model finetuned without any pretraining.}
\label{tab:ablations_pretraining}
}
\end{table}
\subsection{Ablation Studies}\label{sec:pretraining_ablations}
We conduct extensive ablations of our multimodal architecture and cross-modal alignment process, highlighting the contribution of each design choice to overall model performance.

\paragraph{Cross-Modal Alignment} In \autoref{tab:ablations_pretraining}, we find that SALMON pretraining significantly improves both language understanding capabilities and task performance by aligning representations across modalities. Additionally, our STW mechanism provides additional benefits over standard equal token weighting by selectively identifying and emphasizing tokens that benefit most from TS context, leading to more effective cross-modal representation learning.

\begin{table}[htp]
\centering
{\small
\scalebox{0.95}{%
\begin{tabular}{@{}ccc@{}}
\toprule
\textbf{Model} & \textbf{LM Loss} & \textbf{30D} \\ \midrule
Text-Only            & 2.20    &    54.13              \\ \midrule
Shared Parameters           & 2.00    &    55.80              \\
Separate QKV           & 1.85    &    56.78             \\
Separate MLP           & 1.91   &    56.31              \\ \midrule
Cross-Modal Attention, Layers 1-16          & 1.96    &    56.00              \\ 
Cross-Modal Attention, Layers 1-32          & 1.81    &    56.51              \\ \midrule
MSE-ITT            & 1.78    &    57.14              \\
\bottomrule
\end{tabular}
}
\caption{Results demonstrate the value of our proposed MSE-ITT architecture in both language understanding (LM Loss) and financial forecasting (30D) performance, compared to two sets of baselines that (1) share parameters across text and TS inputs and (2) perform cross-modal attention in different layers of the network. Note that the LM Loss results include cross-modal alignment training (SALMON) but the 30D results do NOT.
\label{tab:ablations_architecture}
}}
\end{table}

\paragraph{Modality-Specific Experts}\label{sec:ms_moe} In \autoref{tab:ablations_architecture}, we provide a direct comparison of our method against baseline models that share weights between the text and time-series inputs. These results demonstrate that our multimodal model effectively leverages time series context to improve its language understanding capabilities with unified pretraining (20\% reduction in LM loss). Compared to sharing parameters, our modality-specific experts architecture improves both language and time series understanding capability, mitigating cross-modal interference and allowing unique modality patterns.

\paragraph{Selective Cross-Modal Attention}\label{sec:cma}
We include baselines in which cross-modal attention is performed in early layers of the network, rather than just the second half. As shown in \autoref{tab:ablations_architecture}, cross-modal attention in the early layers of the network is detrimental to performance, supporting our hypothesis that early LM layers focus on low-level features that can become disrupted by premature cross-modal interactions. We believe these findings could generalize to other multimodal contexts (e.g. healthcare, climate, etc.) in which modalities possess different properties but shared temporal alignment.

\section{Model Analysis and Interpretability}\label{sec:model_analysis}
In this section, we explore the behavior of the model and reveal insights into the value of time series context and its potential impact on investment applications.

\subsection{Real-World Case Study}\label{sec:appendix:case_study}
To further illustrate how our model benefits from multimodal reasoning, we examine a real-world event sequence involving the proposed Pfizer, Allergan merger, shown in \autoref{fig:example}. This event unfolded over several months in 2015 and 2016, with headlines describing early merger rumors, confirmation of deal terms, political opposition, and ultimately, the termination of the transaction.

While a headline such as ``Merger terminated by mutual consent'' may appear negative in isolation, it follows a series of prior developments in which investors reacted negatively to news suggesting the deal would proceed. The accompanying stock return history reveals this clearly: mild declines on initial merger rumors, followed by a significant sell-off once the deal was announced, and a strong rebound when the US Treasury issued rules to block the deal. On April 6, 2016, when Pfizer and Allergan officially terminated the merger, the stock surged.

This price behavior indicates that investors viewed the termination as favorable, even though the text of the headline appears quite negative. A text-only model, seeing only the phrase ``merger terminated'', predicts a low probability (0.32) of a positive return. The time series-only model, reacting only to the long-term price trend of the stock price, driven negative by the merger rumors, predicts a score of (0.21). 

However, our multimodal model (MSE-ITT), which jointly reasons over the headline, prior news, and accompanying stock price movements, assigns a high probability (0.79) to a positive return following the announcement. This prediction reflects the model’s ability to perform a form of counterfactual reasoning: although the headline (“merger terminated”) appears negative in isolation, the model recognizes that it cancels a deal investors had previously responded to unfavorably. By conditioning on prior market reactions to similar events, MSE-ITT infers that the termination is likely to be viewed positively. This deep integration of modalities enables the model to capture temporally grounded event semantics, where the sentiment of a headline depends critically on historical context.

Furthermore, this case study illustrates how unimodal models, lacking either narrative or market information, misinterpret semantically ambiguous events. In contrast, MSE-ITT aligns text and time series through token-level weighting that links financial language to observed outcomes, allowing it to resolve sentiment ambiguity and deliver more accurate, context-aware forecasts.

\begin{figure}[t]
  \centering
  \resizebox{\linewidth}{!}{%
  \small
  \setlength{\tabcolsep}{6pt}
  \renewcommand{\arraystretch}{1.25}
  \begin{tabular}{@{}c p{0.46\linewidth} c p{0.28\linewidth}@{}}
    \toprule
    \textbf{Date} & \textbf{News Headline} & \textbf{Stock Return} & \textbf{Interpretation} \\ \midrule
    29 Oct 2015 & \hl{Pfizer approaches Allergan about record merger} & -1.9\%, ... , -0.00\% & Rumor; investors mildly optimistic \\
    23 Nov 2015 & \$160 B deal announced; Pfizer to relocate to Ireland & –2.6\%, ... , +2.3\% & Fear of dilution and tax-inversion risk \\
     5 Apr 2016 & U.S. Treasury issues rules likely to scuttle tax inversion & +2.1\% & Hope deal will be abandoned \\
     6 Apr 2016 & Merger \textbf{terminated} by mutual consent & +5.0\% & Relief rally \\
    \bottomrule
  \end{tabular}}
  \caption{Example sequence of news events and market reactions about the potential Pfizer-Allergan merger. The news articles contain persuasive language, yet the accompanying stock returns reveal the market’s true perception: optimism on early rumors, a sharp sell-off once the deal terms are set, and a relief rally when the takeover is cancelled. Jointly modeling these inputs can more effectively interpret the outcome of such events.}
  \label{fig:example}
\end{figure}

\subsection{Value of Time Series Context}\label{sec:value_ts_context} To further investigate where and when time-series context is most beneficial, we use the LM Financial Dictionary \citep{loughran2011liability} to classify the financial sentiment of words. Following \autoref{sec:salient_pretraining}, we compute the benefit of TS-context (likelihood ratio) for each group of words across the test set with our MSE-ITT model after SALMON pretraining.

\begin{table}[t]
\centering
\scriptsize
\setlength{\tabcolsep}{5pt}
\begin{tabular}{l c}
\toprule
\textbf{Word Category} & \textbf{Likelihood Ratio} \\
\midrule
Stop Words       & 0.71 \\
Non-Sentiment    & 1.45 \\
All-Sentiment    & 1.83 \\
Positive         & 2.17 \\
Negative         & 1.74 \\
Litigious        & 1.76 \\
Uncertainty      & 1.63 \\
Constraining     & 1.23 \\
Weak Modal       & 2.95 \\
Strong Modal     & 1.28 \\
\bottomrule
\end{tabular}
\caption{Median likelihood ratios across word categories from the LM Financial Dictionary \citep{loughran2011liability}, computed on the test set. These ratios quantify the marginal benefit of time series context for predicting words in each category, revealing the strongest gains for sentiment-charged words.}
\label{tab:lm_dict}
\end{table}

In \autoref{tab:lm_dict}, we report median statistics on the likelihood ratio for sentiment words, as defined by the LM Financial Dictionary \citep{loughran2011liability}, which indicates the marginal benefit that time series context provides for predicting the identity of the tokens. These results demonstrate that sentiment-charged words benefit from the time series context significantly more than non-sentiment words, and dramatically more than stop words. These findings highlight the complementary nature of time series and textual data and reinforce the design of our \textbf{STW} loss function \autoref{sec:salient_pretraining}. The time series context provides valuable information for improving language understanding and interpreting the impact of news events.

\subsection{Portfolio Simulations}\label{sec:port_sims}
In \autoref{tab:port_sims}, we demonstrate the economic value of our methodology with portfolio simulations. We form monthly long-short (\textit{market-neutral}) portfolios \citep{fama2015five} by sorting stocks based on the 30D model predictions from the past month, detailed in \autoref{sec:appendix:port_sims}. For comparison, we simulate unimodal baselines and the best performing multimodal baselines, described further in \autoref{sec:appendix:port_sims}. Following \citet{cong2021alphaportfolio}, we include net performance that includes conservative estimates of the impact of transaction costs on portfolio implementation. 

The test period (Jan 2020 - Dec 2024) spans 5-years of highly diverse market regimes, including the COVID-19 crash and recovery, stimulus-driven expansion, and inflation and rate hikes, ensuring robustness across economic conditions. The resulting performance of our unified multimodal model generates investment performance that is economically and statistically better than the best performing multimodal baselines. These results demonstrate that our model provides significant predictive value in a real-world trading setting. 

\begin{table}[htp]
\centering
{\small
\scalebox{0.70}{%
\begin{tabular}{@{}cccc@{}}
\toprule
\textbf{Method} & \textbf{Net Return} & \textbf{Volatility} & \textbf{\begin{tabular}[c]{@{}c@{}}\textbf{Net Sharpe Ratio}\end{tabular}} \\ \midrule
TS-Only \citep{nie2022time} & 5.99 & 13.11 & 0.46  \\
Text-Only \citep{chen2022expected} & 8.60 & 10.47 & 0.82  \\
TTSR \citep{Chow2024TimeSeriesReasoning} & 12.37 & 11.28 & 1.10  \\
Hybrid-MMF \citep{Kim2024MMF} & 11.60 & 10.19 & 1.13  \\
MTFE-MICM \citep{koval-etal-2024-financial} & 10.23 & 10.39 & 0.99  \\
StockTime \citep{wang2024stocktime} & 13.91 & 12.65 & 1.10  \\
\textbf{Proposed, MSE-ITT} & \textbf{17.01} & \textbf{11.26} & \textbf{1.51} \\ \bottomrule
\end{tabular}} 
} 
\caption{Annualized portfolio statistics of simulated investment performance, expressed in percentage units. ``Net'' performance includes an estimate of the impact of transaction costs, detailed in \autoref{sec:appendix:port_sims}.}
\label{tab:port_sims}
\end{table}

\section{Conclusion}
We propose a unified multimodal architecture for modeling interleaved sequences of text and time series data, and introduce a cross-modal alignment framework with a salient token weighting mechanism that learns to align representations across modalities with a focus on the most informative tokens. Our approach demonstrates state-of-the-art performance on a challenging financial forecasting task, and our ablation experiments confirm the contribution of our design decisions. These forecasting improvements translate to economically meaningful gains in portfolio simulations, underscoring the real-world value of our approach. These findings highlight the need for modality-specific structures and joint reasoning in multimodal LMs, with broader implications for domains in which text and time series co-evolve.

\paragraph{Limitations}
Our findings demonstrate that large language models can benefit significantly from structured time series context when modeling interleaved sequences of text and numerical data, but only with the appropriate inductive biases. The proposed architecture improves both forecasting accuracy and simulated investment performance through modality-specific routing and specialized cross-modal pretraining. However, our experiments focus on financial forecasting within English financial news and US-based stocks, but we believe the methodology is applicable to other domains involving interleaved text and time series (e.g. healthcare, climate), which we leave to future work. While we incorporate conservative estimates of transaction costs in our investment simulations, real-world trading requires more detailed consideration of trade execution and risk management, which we leave to future work. Please note that our financial prediction system is intended for research use and that portfolio results are presented for illustration purposes only, not as investment advice.

\bibliography{custom}

\appendix
\section{Appendix}
\label{sec:appendix}

\subsection{Pretrained Language Models}
We implement all models in PyTorch and source all pretrained checkpoints from HuggingFace. 

\subsection{Baseline Models}\label{sec:appendix:baseline_models}
For our proposed model (MSE-ITT), for supervised classification, we train a classification head on top of the end-of-sequence token's last hidden representations to make binary predictions.  

For some baseline models in which additional inputs or modalities are incorporated, such as proprietary sentiment scores \citep{Wang2024ModelingNews}, we do not include these additional inputs (and we do not have access to them) in the model in order to present a fair comparison of model types and isolate the effects of multimodal fusion between sequences of text and time series. Some baseline models \citep{Wang2024ModelingNews, Liu2024TimeMMD, Li2025TaTS} explore the use of various LLMs for textual encoding. In such cases, we select the LM with the best reported performance for evaluation. For models that patch the time series input into non-overlapping consecutive chunks \citep{Chow2024TimeSeriesReasoning, wang2024stocktime}, if the patch length used is not provided in the implementation details, then we tune the value over $\{1, 5, 10\}$ based on validation set performance. Further, for some multivariate time series baseline models that require a one-to-one mapping between text and time series inputs across time steps \citep{koval-etal-2024-financial, Li2025TaTS}, if there are no news articles on a given time step, then we simply carry forward the previous news article (embeddings) until the next news article (embeddings) are available. Additionally, in the MAT baseline \citep{emami2024modality}, the authors do not report the number of topics or the base pretrained LM used in BERTTopic \citep{grootendorst2022bertopic}, so we resort to using the hidden representations produced from their sentiment model FinBERT \citep{araci2019finbert} as text features. Further, for the FinMA \citep{xie2023pixiu} baseline, we provide the text and time series sequences in the same format as for GPT-4o \autoref{sec:appendix:zeroshot_llms}, such that the time series is converted to raw text and interleaved between news articles in temporal order.

For our proposed model, we use learnable special tokens to denote the beginning and end of each news article and modality \citep{caciularu2021cdlm}, and include the article timestamp in the article text.

\subsection{Portfolio Simulations}\label{sec:appendix:port_sims}
In \autoref{tab:port_sims}, we demonstrate the economic value of our model predictions using portfolio simulations. 

To perform these simulations in a realistic setting, we first filter the investment universe of stocks according to sufficient liquidity requirements to ensure feasibility of portfolio implementation, including a minimum market capitalization of \$250M and daily average value of shares traded of \$1M. 

Then, we form monthly long-short (market-neutral) quintile portfolios according to \citet{fama2015five}. To this end, we sort stocks based on the average 30D model predictions from news articles about companies in the past 1 month. 
Then our portfolios are formed by buying those in the top 20\% of average scores and shorting those in the bottom 20\% of average scores on a monthly basis in equal proportions. Please note that these portfolios are market-neutral and therefore have essentially no correlation with broad market indices. 

In \autoref{tab:port_sims}, we include conservative estimates of the impact of transactions costs on portfolio implementation. We follow the turnover-based method used in \citet{cong2021alphaportfolio}, which conservatively estimates the annual transaction cost as 0.01 times the annual 1-way portfolio turnover. Therefore, the net return of the portfolio is the gross return minus the estimated transaction costs. 

\subsection{Data Curation}\label{sec:appendix:data_curation}
The FNSPID dataset spans multiple news sources, including Nasdaq, Reuters, CNBC, Benzinga, and cover a variety of company events, including product launches, earnings reports, and mergers. We truncate each article after the first 128 words during all experiments for computational efficiency. We only include articles originally written in English according to the following criteria \citep{chen2022expected}: they are tagged relevant for only one company; they are longer than 100 characters or shorter than 10,000 characters; they contain less than 10\% of numerical characters; they have less than 90\% Jaccard similarity to a previous article (to remove potential duplicate articles). We require that each company possess at least 5 news articles within the past year as well as available stock returns to be in our sample. In addition, we apply further quality filtering to ensure that the dataset contains high-quality, event-driven news articles that are specific to individual stocks. We systematically remove broad market summaries, sector-level reports, and generic financial commentary by filtering out headlines containing specific keywords and patterns. This process allows us to isolate stock-specific events. There is a slight class imbalance so we randomly downsample the majority class to ensure balanced class ratios. 

\subsection{Zero-Shot LLMs}\label{sec:appendix:zeroshot_llms}
For the zero-shot GPT-4o baselines, we have explored a variety of prompting strategies, listed below. For this set of baselines, the time series inputs are converted to text strings and we tune their form \{decimal, percentage\}, digits of precision \{2, 3, 4\}, and whether to include a space in between digits, with validation set performance \citep{gruver2023large, Williams2024ContextIsKey}. We do not find much variation in the performance of each, but tune them according to the validation set and report those results in \autoref{tab:main_results}. 

To this end, we construct a natural language prompt that interleaves company-specific financial news with daily stock return sequences in chronological order. The model is tasked with predicting future price movement on a bounded scale.

Each prompt follows the format:

\begin{quote}
\texttt{Given the company's daily stock return over the last year interleaved in temporal order with recent news about the company, sorted chronologically:}\\
\texttt{\{interleaved\_input\}}\\
\texttt{Predict the company's future stock price movement over the next \{horizon\} days, on a scale from 0 to 100, where 100 indicates strong positive movement and 0 indicates strong negative movement. \{prompt\_style\}}\\
\texttt{Prediction:}
\end{quote}

The field \texttt{\{interleaved\_input\}} contains alternating sequences of news articles and their corresponding stock returns (e.g., \texttt{News Article: ... Stock Returns: ...}). The field \texttt{\{prompt\_style\}} controls the reasoning strategy used by the model and takes one of two forms:

\begin{itemize}
    \item \textbf{Direct}: The model immediately outputs a prediction.
    \item \textbf{Chain-of-Thought (CoT)}: The model is instructed to ``think step-by-step'' before producing a prediction, following recent findings that CoT improves LLM reasoning over time series data~\citep{Tan2025InferringEvents}.
\end{itemize}

This design enables a uniform comparison of reasoning strategies across interleaved multimodal inputs, without any task-specific fine-tuning.

\subsection{Statistical Significance}\label{sec:stats}
In \autoref{tab:main_results}, we report the sample standard deviation of results for our proposed method across 3 different training runs with different random seeds. The variability of results across random seeds stems from the randomness in the training process caused by random initialization of the classification layer weights and the random batch order of training samples during stochastic gradient descent optimization. We also report the results from DeLong's pairwise test of statistical significance between model AUC scores \citep{delong1988comparing}.

\subsection{Time Series Discretization}\label{sec:appendix:ts_quant}
For time-series discretization, we tune the number of discrete time-series bins $B\in \{4, 8, 16, 32, 64\}$ and their embedding dimension $d_{ts} \in \{32, 64, 128, 256, 512\}$. To map the time-series embeddings into the token embedding space of the LM, we learn a simple linear map to increase dimensionality from $d_\text{ts}$ to $d_\text{text}$. 

In \autoref{tab:appendix:ablations:ts_quant}, we compare our discretization approach to alternative continuous embedding methods. While the benefits of discretization are indeed modest and clearly not the primary source of our performance gains, they dramatically simplify and unify the interleaved multimodal (SALMON) pretraining objective. By converting all tokens, regardless of modality, to a common discrete space, we can apply a consistent cross-entropy loss to all discrete tokens, without inducing a complex multitask setup that requires careful calibration of continuous and discrete loss terms. In the continuous setting, we tune the weight on the TS (Mean Squared Error) loss over $w_{ts} \in \{0, 10, 100, 1000\}$ according to validation set performance. 

\begin{table}[htp]
\centering
{\small
\scalebox{0.90}{%
\begin{tabular}{@{}cccc@{}}
\toprule
\textbf{Method} & \textbf{LM Loss} & \textbf{7D} & \textbf{30D} \\ \midrule
Discrete & 1.78 & 57.94 & 58.48 \\ 
Linear   & 1.83 & 57.54 & 58.10 \\
MLP      & 1.83 & 57.62 & 58.13 \\
\bottomrule
\end{tabular}} 
} 
\caption{Results demonstrate the value of time series discretization in our multimodal architecture in both language understanding (LM Loss) and financial forecasting (7D, 30D) performance, compared to baselines that use continuous embeddings via linear and nonlinear transformations. These results include SALMON pretraining.}
\label{tab:appendix:ablations:ts_quant}
\end{table}

\subsection{Training Details and Hyperparameter Tuning}\label{sec:appendix:implementation}
We perform all experiments on a single NVIDIA H100 GPU with 80G in memory. We use AdamW to optimize all parameters. For all finetuned models, we use an effective batch size of 64 with gradient accumulation. We train all models for up to 5 epochs based on validation set performance for test evaluation. All supervised models are optimized using Binary Cross-Entropy as the loss function.

We tune the learning rate over \{1e-6, 3e-6, 5e-6, 7e-6, 1e-5\} and LoRA \citep{hu2021lora} rank parameter over $r\in \{16, 32, 64\}$ according to validation set performance. We apply LoRA adapters to all linear layers. We use a simple 2-layer MLP classification layer to project the token hidden states to make a binary prediction for all language models used.

For computational constraints, we train all models using mixed precision training and gradient checkpointing to satisfy GPU memory constraints, and clip gradient norms. For Llama-based models, we finetune in BF16 precision. For LoRA-based finetuning, we always set the value of the alpha parameter to be equal to double the value of rank parameter. 

\subsection{Multivariate Time Series}\label{sec:appendix:multivariate}
In this section, we demonstrate that our MSE-ITT model architecture can support numerical time series with multiple channels. 

To this end, we select 15 commonly used market price and accounting-based financial variables available at the time of the report date from the definitions and cluster classifications in \citet{swade2023factor}. This set includes dividend yield (Value), earnings-to-price (Value), sales-to-price (Value), book value-to-price (Value), sales growth (Growth), earnings growth (Growth), gross profit to assets (Profitability), net income to equity (Profitability), net income to assets (Profitability), medium-term price momentum (Momentum), short-term price reversal (Reversal), price volatility (Low Risk), market leverage (Debt Issuance), share turnover (Low Risk), and market capitalization (Size). 

Since some variables have different frequencies, ranging from daily to quarterly, we up-sample them all to daily frequency by forward-filling previous values.  

We fit different discretized embeddings for each channel following our approach in \autoref{sec:methods}. After this, we simply concatenate their embeddings together vertically at a daily frequency and similarly interleave with the news articles according to timestamp. To extend our SALMON pretraining objective to the multivariate setting, we learn a separate output projection head for each channel and compute the discretized token for each channel independently and average the loss across channels. 

\begin{table}[htp]
\centering
{\small
\scalebox{0.90}{%
\begin{tabular}{@{}ccc@{}}
\toprule
\textbf{Method} & \textbf{LM Loss} & \textbf{30D} \\ \midrule
Univariate              & --   & 57.14 \\ 
Univariate w/ SALMON    & 1.78 & 58.48 \\
Multivariate            & --   & 58.45 \\
Multivariate w/ SALMON  & 1.74 & 59.89 \\
\bottomrule
\end{tabular}} 
} 
\caption{Results indicate the performance of our MSE-ITT multimodal architecture when extended to the multivariate setting in both language understanding (LM Loss) and financial forecasting (30D) performance.}
\label{tab:appendix:multivariate}
\end{table}

\end{document}